\documentclass[10pt,floatfix]{article}

\usepackage{graphics}
\newcommand{\bea}{\begin{eqnarray}}
\newcommand{\eea}{\end{eqnarray}}
\newcommand{\be}{\begin{equation}}
\newcommand{\ee}{\end{equation}}

\def\be{\begin{eqnarray}}
\def\ee{\end{eqnarray}}
\def\bd{\begin{displaymath}}
\def\ed{\end{displaymath}}

\def\etal{{\em et al.}}

\def\NP{Nucl. Phys. }
\def\PR{Phys. Rev. }
\def\PRL{Phys. Rev. Lett. }
\def\PL{Phys. Lett. }
\def\jpg{J. Phys. G: Nucl. Part. Phys. }

\begin{document}
\title{Helium nuclei around the neutron drip line}
\author{Madhubrata Bhattacharya and G. Gangopadhyay\\
Department of Physics, University of Calcutta\\ 92 Acharya Prafulla 
Chandra Road, Kolkata-700 009, India\\
Subinit Roy\\
Saha Institute of Nuclear Physics\\ Block AF, Sector 1, 
Kolkata- 700 064, India}
\maketitle

\begin{abstract}
Neutron rich He nuclei have been investigated using relativistic mean field 
approach in co-ordinate space. Elastic partial scattering cross sections for 
proton scattering in inverse kinematics have been calculated using the 
theoretically obtained density for $^{6,8}$He and compared with experiment. 
The energies of the low-lying resonance states in the neutron unstable nuclei 
$^{5,7}$He have also been calculated and compared with experimental 
observations.

\end{abstract}

\section{Introduction}

Improvement in experimental techniques in the last decades has led to 
the production and study of very light neutron rich nuclei up to and 
even beyond the neutron dripline. One of the very interesting phenomena 
in such nuclei is the neutron halo\cite{exotic}. The halo significantly 
affects different reactions involving these nuclei.

In an earlier work\cite{CBe}, we studied the structure of exotic even-even 
Be and C nuclei and calculated the elastic proton scattering cross section 
using the theoretical densities. In the present work, we apply the same 
procedure  
to He nuclei near the neutron drip line. 
These nuclei have only a few nucleons and show a very 
large neutron-proton ratio. Study of such nuclei is important for
the effect of their extreme isospin values on the nuclear interaction. 
Besides the bound states in the even-even nuclei, low energy resonance states 
in odd mass He nuclei beyond the drip line have also been 
investigated. 

One of our main interest lies in the prediction for neutron radius and neutron 
density in He nuclei. Neutron rich He nuclei are known to exhibit neutron 
halos.  However, determination of the extent of the halo 
is ambiguous, as the information on density is model dependent in absence of 
direct measurements like electron scattering. For a nucleus with only a few 
nucleons, the bulk radius value, extracted from experiment, may also be model 
dependent. Direct comparison with experimental measurements may yield better
idea about the accuracy of the calculation. For example, calculation of 
differential
cross section in elastic proton scattering in inverse kinematics is expected
to provide a test for the calculated densities\cite{EP}. 

\section{Method}

Relativistic mean field (RMF) approach is now a standard tool in low energy
nuclear structure. 
It has been able to explain different features 
of stable and exotic nuclei like ground state binding energy, deformation, 
radius, excited states, spin-orbit splitting, neutron halo, etc\cite{RMF1}. 
It is well known that in nuclei far away from the 
stability valley, the single particle level structure undergoes certain 
changes in which the spin-orbit splitting plays an important role.   
RMF is particularly suited to investigate these nuclei because it 
is based on the Dirac Lagrangian density which naturally incorporates the
spin degrees of freedom.

Different variations of the Lagrangian density, as
well as different parametrizations, have been systematically investigated
by many workers. 
In our earlier work\cite{CBe}
we used the density NLSH\cite{NLSH}, known for its ability to describe 
nuclei near stability valley. A newer Lagrangian
density, FSU Gold, which involves self-coupling of the vector-isoscalar 
meson as well as coupling between the vector-isoscalar meson and the 
vector-isovector meson, was proposed in Ref.\cite{prl}. 
This density was applied in our studies on proton 
radioactivity\cite{plb}, cluster decay\cite{prc1} and alpha 
decay\cite{prc2}, etc. 
NL3\cite{NL3} is another force that has proved to be very useful in describing 
the ground state properties throughout the periodic table. Another 
force, NL2\cite{NL2}, has been found to be successful for the
description of light nuclei.
In the present work, we employ all the above forces and compare the results.  

In the conventional RMF+BCS approach for even-even nuclei, the Euler-Lagrange 
equations are solved under the assumptions of classical meson
fields, time reversal symmetry, no-sea contribution, etc. Pairing is introduced
under the BCS approximation. 
Usually the resulting equations are solved in a harmonic oscillator basis.
However, in exotic nuclei, the basis expansion method using 
harmonic oscillator, because of its incorrect asymptotic properties, face 
problems in describing the loosely bound halo states. 
A solution of the Dirac and Klein Gordon equations in co-ordinate space may be 
preferable to describe the weakly bound states. 
Such calculations exist in Relativistic Hartree-Bogoliubov (RHB) approximation 
in r-space. 

The RHB calculations, or their nonrelativistic counterparts  
Hartree-Fock-Bogoliubov equations are very involved and time consuming. 
Interested readers are referred to the calculation of \cite{cpc}. Particularly 
important are the Relativistic continuum Hartree-Bogoliubov calculations
\cite{rchb}
which take the continuum into account.
A simpler approximation, introduced in Refs\cite{conpair,hfb1,hfb2},
takes into account the effect of the resonant continuum through the scattering 
wave functions located in the region of the resonant states in the 
nonrelativistic picture. 
The cases of zero range and finite range pairings have also been investigated 
and the importance of the truncation of the quasiparticle space for zero range 
pairing interaction has been highlighted in the above works. 

RMF equations involving continuum states have also been
solved\cite{rmfc,rmfc1} with scattering wave functions. All these calculations
have taken into account the width of the continuum levels also. For example,
Cao and Ma \cite{rmfc1} have also compared their results with those from
calculations
with zero width. They conclude that the pairing gaps, the Fermi levels, the
pairing correlation energies, and the binding energies are considerably affected
by proper consideration of the width of the resonant states. Particularly,
near the neutron drip line, this effect is expected to be very important.

We have used the above co-ordinate space RMF+BCS approach 
earlier\cite{CBe,CaNi} to study neutron rich nuclei in different mass regions. 
The proximity of the neutron rich nuclei to the drip line necessitates one to 
consider the effect of the positive energy states. The widths of the positive 
energy levels have been taken into account in the present work.  
We have
confined out calculation to spherical approximation as He nuclei are expected
to be spherical.  Odd nuclei have been investigated in the tagged approximation.

For the solution of the equations in co-ordinate space, the mesh size has been 
taken as 0.04 fm. We assume the nuclear interaction to
vanish at a radius of 14 fm. We have checked that  an increase in the 
last quantity to 20 fm keeps the results almost unchanged. The total energy 
varies by less than 0.02\%, and the neutron radius, by less than 1\%.  
Usually, the strength of the pairing interaction is chosen to reproduce the 
pairing energy in RHB calculations\cite{rmfc}. However,  in our calculation
the strength of the 
zero range volume pairing force is taken as 400 MeV-fm$^3$ for neutrons from 
systematics as it was found to explain the trend in binding 
energy in very light nuclei reasonably well.
For example, in Table \ref{be} we list the FSU Gold results for binding energy 
in a few very light nuclei. For Be nuclei, the proton pairing  strength was 
taken as 200 MeV-fm$^3$. It should be mentioned that the result for $^{10}$Be 
is for a deformed calculation in the method followed in \cite{rmf,CBe}. 
In Ref. \cite{CBe}, we found $^{10}$Be to be strongly deformed.
 It has been also observed that 
changes of the order of 10\% in the value of the neutron 
pairing strength do not affect our conclusions appreciably.
\begin{table}
\caption{Binding energy per nucleon in a few very light nuclei 
using FSU Gold Lagrangian.
Pairing strength for protons and neutrons have been taken as 200 MeV-fm$^3$
and 400 MeV-fm$^3$, respectively.\label{be} See text for details.}
\begin{tabular}{rcc}\hline
Nucleus & \multicolumn{2}{c}{Binding Energy (MeV)}\\
 & Expt. & Theo.\\\hline
$^{7}$Li & 5.606 & 5.535\\ 
$^{9}$Li & 5.038& 5.121 \\
$^{10}$Be & 6.498 & 6.636\\
$^{12}$Be & 5.721 & 5.696\\
$^{14}$Be & 4.994 & 4.854\\
\hline
\end{tabular}
\end{table}

As pointed out by Sandulescu \etal\cite{rmfc}, in the RHB equations the 
pairing cut off is usually very large allowing the quasiparticles to 
scatter over a very large energy. In contrast, in the RMF+BCS calculations,
only a few resonant states around zero energy are included. In our case, we
included only the states up to the $p$-shell in the rmf calculation. Thus the 
maximum quasiparticle energy corresponded to $1s_{1/2}$ state, and the
cut-off, to approximately 25 MeV.

Electron scattering, the most direct method for measuring nuclear density, is 
difficult to apply far away from the valley of stability. Elastic proton 
scattering in inverse kinematics provides an alternate test for the calculated 
densities\cite{EP}. 
The optical model potential is obtained using an effective interaction,
derived from nuclear matter calculation, in the local density approximation,
{\em i.e.} by substituting the nuclear matter density
with the calculated density distribution of the finite nucleus.  
In the present case microscopic nuclear potentials have been obtained by folding two 
effective interactions, discussed later, with the microscopic densities obtained in the RMF 
calculations. The Coulomb potential has been similarly obtained by folding the 
Coulomb interaction with the microscopic proton density. 

A common effective interaction DDM3Y\cite{ddm3y1,ddm3y2}
was obtained from a finite range energy independent M3Y interaction by adding
a zero range energy dependent pseudopotential and introducing a density dependent factor.
This
interaction was employed widely in the study of nucleon nucleus as well
as nucleus nucleus scattering, calculation of proton radioactivity, etc.
The density dependence may be chosen as exponential\cite{ddm3y1} or be
of the form $C(1-\beta\rho^{2/3})$\cite{ddm3y2}. In this particular work we
have selected the latter form.
The constants, obtained from nuclear matter calculation\cite{ddm3y3}
as $C=2.07$ and $\beta=1.624$ fm$^2$, have been used in our
calculation. For scattering we have taken
real and the imaginary parts of the potential as 0.8 times and 0.2 times
the DDM3Y potential. In both the calculations, the spin-orbit potential was
chosen from the Scheerbaum prescription\cite{SO}.
The calculations have been
performed with the computer codes MOMCS\cite{MOMCS} and ECIS95\cite{ECIS}
assuming spherical symmetry.

To check our results, we employed another interaction, 
the JLM interaction of Jeukenne, Lejeune, 
and Mahaux (JLM)\cite{JLM74}  in which further 
improvement was incorporated in terms of the finite range of the effective
interaction by including a Gaussian form factor. 
We have used the global parameters for the 
effective interaction and the respective default normalizations for the 
potential components from Refs. \cite{MOMCS} and \cite{MOM} with Gaussian 
range values of $t_{real}=t_{imag}=1.2$ fm. No search has been performed on 
any of these parameters. 

\section{Results}

\subsection{Even-even isotopes - ground state energy and density}

The two neutron rich even-even He nuclei, stable against neutron emission, are
$^{6,8}$He. Experimental measurements exist for binding energy and radius 
values in these nuclei.  The latter have been measured in different ways. 
Using proton elastic scattering Kiselev \etal 
\cite{He6rad} measured the matter radii in $^{6,8}$He to be 2.37(5) fm and 
2.49(4) fm, respectively. 
Tanihata \etal\cite{Tan} obtained the matter radii of  $^{6,8}$He as 2.33(4) 
and 2.49(4) fm, respectively. A re-analysis\cite{Alkhalili} of the same data
yielded the value 2.71(4) fm for $^6$He. Lapoux \etal ~found a radius of 2.5 
fm from inelastic scattering data\cite{Lapoux1} and 2.55 fm from elastic
scattering\cite{Lapoux2} for $^6$He. Finally, Glauber model analysis in the 
optical limit yielded the values 2.48(3) fm and 2.52(3) fm\cite{OST} for 
$^{6,8}$He, respectively. In another experiment, 
Egelhof \etal\cite{Egelhof} 
deduced the values to be 2.30(7) fm and 2.45(7) fm, respectively,  from
proton scattering at intermediate energy. Using the proton radius value
quoted in Table \ref{tab1}, measured by laser spectroscopy\cite{pHe6}, the 
experimental neutron radii are seen to lie within the range 2.41 - 2.98 fm for
$^6$He and 2.60- 2.69 fm for $^8$He.

\begin{table}
\caption{Binding energy per nucleon and radius values in $^{6,8}$He
compared with experimental values. Experimental 
binding energy values are from the compilation\cite{AWT}. Experimental
proton radii are from laser spectroscopy study\cite{pHe6}.
Experimental r.m.s. radii values are the results of Glauber model analysis in 
the optical limit\cite{OST}. 
See text for radius values from other measurements. \label{tab1}}
\center
\begin{tabular}{cccclll}\hline
&&$^{6}$He & $^{8}$He & \\\hline
         &Expt. &4.878 &3.926\\
B.E.(MeV)&NLSH&5.831&4.669\\
         &FSU Gold&5.507&4.222\\
         &NL3 & 5.890 & 4.780  \\
         &NL2  &5.161 & 3.765 \\
\hline
$r_p$(fm)  
         & Expt. & 2.068(11)&1.929(26)\\
         &NLSH&1.86&1.83&\\
         &FSU Gold&1.88&1.86\\
         &NL3 & 1.92 & 1.88 \\
         &NL2  & 1.89& 1.88\\\hline
$r_n$(fm) &  Expt. & 2.72 & \\
          &NLSH&2.92&2.83&\\
          &FSU Gold&3.12&3.07\\
          &NL3 &  3.11 &2.89\\
          &NL2  & 3.63  &3.69\\
\hline
$r_{rms}$(fm)  
         &Expt.&2.48(3)&2.52(3)&\\
         &NLSH&2.61&2.61&\\
         &FSU Gold&2.76&2.81\\
         &NL3 & 2.77&2.67  \\
         &NL2  &3.16&3.33   \\\hline
\end{tabular}

\end{table}
In Table \ref{tab1}, our results for binding energy and radius values in 
$^{6,8}$He are given and compared with experimental measurements 
wherever available. There is a basic difference between 
the single particle levels predicted by the different forces. The two forces 
NL2 and FSU Gold, which produce better agreements with the binding energy 
values, predict the level $\nu p_{1/2}$ to be in the continuum in $^8$He. The 
other two forces predict it to be very weakly bound. We find that NL2 
predicts both $^{6,8}$He to have a negative two neutron separation energy at 
variance with experimental observations.

As already mentioned, the effect of the width of the levels in the continuum 
has been incorporated in our calculation. For example, FSU Gold predicts the 
level $\nu p_{1/2}$ to be in the continuum in $^6$He with a very large with 
width 1.8 MeV, which, in $^{8}$He, comes down to 0.28 MeV. Obviously, the 
angular momentum of the $p_{1/2}$-state being small, the centrifugal barrier 
cannot localize the state very effectively. The effect of the resonant level 
on the binding energy is very small (less than 0.05\% in the case of 
$^6$He) as the occupancy of $p_{1/2}$ level is very small. However, even this
small occupancy has a larger effect on the neutron radius concerned, because 
an unbound resonant level has a radius much larger than the bound state. Thus, 
we find that the effect of including the effect of the level width increases 
the neutron radius by 0.5\%. Small though the number is, it is comparable to
the experimental errors in measurements of rms radii in these nuclei. 

The experimental radii values that have been shown in Table \ref{tab1} are 
from Glauber model analysis\cite{OST} and laser spectroscopy\cite{pHe6}. 
The calculated results are in reasonable agreement with experimental 
measurements given the fact that the number of nucleons is very small and the 
mean field approach may not be very accurate. One can see that the force NL2,
which gives the least binding energy, also predicts neutron radii
values to be considerably larger than the the other forces.

The calculated proton and the neutron densities in $^{6,8}$He are shown in 
Fig. 1. One interesting observation is that the at large radius, density in 
$^{6}$He decreases more slowly than in $^{8}$He. This is the reason that the 
neutron radius of the latter nucleus is smaller than that of the former.
This is obviously due to the fact that the level $\nu p_{1/2}$ is either bound 
or have very small positive energy in $^8$He. The two forces, NLSH and NL3, 
which predict this level to be bound in $^8$He, show a smaller neutron radius 
as expected.
\begin{figure}
\vskip -2cm
\resizebox{9cm}{!}{\includegraphics{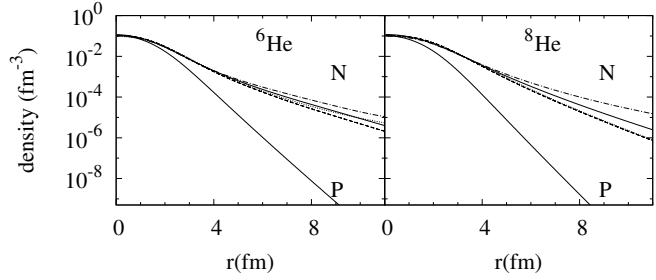}}
\caption{Calculated proton and neutron densities in $^{6,8}$He for
the  force FSU Gold (solid line), NLSH (dashed line), NL3(dotted line), and
NL2 (dash-dotteded line). The proton density is nearly identical in all the 
calculations.}
\end{figure}

\begin{figure*}
\resizebox{10cm}{!}{\includegraphics{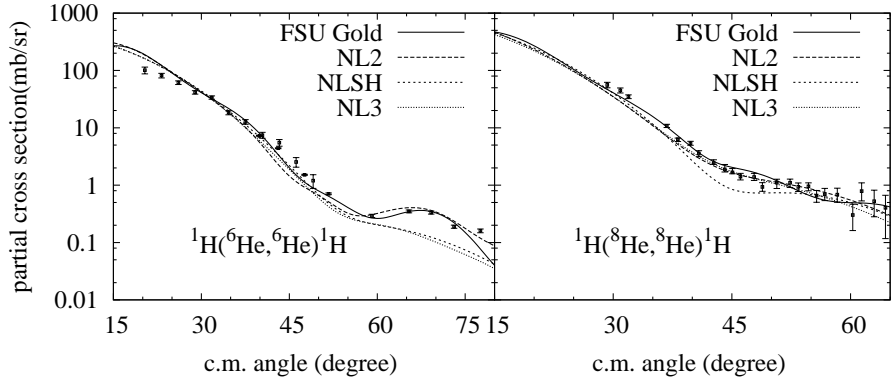}}
\caption{Partial cross section for the elastic proton scattering 
in inverse kinematics using the DDM3Y interaction. The projectile 
energies of $^{6}$He and $^8$He are 71 MeV/A and 72 MeV/A, respectively.}
\end{figure*}
In view of the ambiguity of the radius extracted from different measurements, 
we have calculated the differential cross sections directly using two standard 
interactions, JLM and DDM3Y, for some of the actual experiments. All the 
densities obtained from different Lagrangians were used.
In Fig. 2
we select the experimental values for $^{6,8}$He scattering from Refs. 
\cite{EL,EL1} and \cite{EL2}, respectively and compare with our calculation. 

As we have seen, the nuclear densities from NLSH, NL3 and FSU Gold are 
very similar, while NL2 shows a more diffused neutron density and larger 
neutron radius. However, we find that in the region where experimental data is 
available, there is very 
little difference between the predictions of the forces. The force FSU Gold 
shows a larger cross section at very large angles, a region where data 
is not yet available. The DDM3Y interaction describes the data much better, 
particularly at large angles. The results of the JLM interaction (not shown in 
the figure) show a smoother behaviour maintaining the overall trend. 
In $^6$He, the results for NL3 and NLSH densities can be brought closer to 
experiments by modifying
parameters such as normalization factors of the potential or the ranges
in the Gaussian form factors in the MOM approach. However, no attempt
has been made to fit the data and global parameters have been adopted in the 
present work. 
From these results, it is possible to
conclude that the density has been predicted reasonably well in 
the present calculation. Results for other energies show similar 
agreement. 

\subsection{Odd isotopes - resonance states}

As shown above, the present method provides a reasonable description of 
the even-even nuclei $^{6,8}$He. We note that these nuclei are stabilized
against neutron emission by the pairing force. We next extend our study to
the odd nuclei $^{5,7}$He which are unstable against neutron 
emission. 

The most important experimental information in these odd isotopes are the 
energies of some of the states in the continuum. It is possible 
to calculate the energy of some pure single particle resonances and compare 
with experiment. Though the total energy is not very accurately predicted, we 
expect the resonance energy, being the difference of two absolute energy 
values, to be more accurate. Thus, it is possible to probe the structure of 
nuclei beyond the neutron drip line. We study the one quasiparticle resonances 
built on the single particle states $1p_{3/2}$ and $1p_{1/2}$. They correspond 
to the observed states built on the ground state of the even-even core plus 
the last odd neutron in the single particle orbits mentioned above.

As we will see, the experimental situation is rather unclear in these nuclei.
The measurements about which there are some degree of agreement between 
different experiments are the energies of the $3/2^-$ resonance.
We find that the FSU Gold Lagrangian density provides the best overall results 
for these values. The NLSH results are comparable but slightly poorer. The 
results from NL2 are much worse while NL3 predicts $^7$He to be 
stable against neutron emission.
For odd mass He nuclei, we present the results for the FSU Gold density only. 

To very briefly summarize the experimental situation in $^5$He, the 
lowest energy states are known to arise out of ground state of the even-even 
core coupled to a $p_{3/2}$ neutron. In $^{5}$He, this state occurs at a 
resonance energy around 0.8 MeV. Tilley \etal\cite{Til} placed it at 0.798 MeV
with a width 0.648 MeV and found another resonance 
with spin-parity  $1/2^-$ at 2.068 MeV with width 5.57 MeV. Although the 
results for the ground state resonance was in reasonable agreement with earlier
measurements\cite{Sal}, the situation in  $1/2^-$ was different. Here, 
previous work placed the resonance at 4.089 MeV. In a recent analysis
\cite{Aksyutina} of an older work\cite{Mark}, the ground state was found 
at a resonance energy of 0.741(4) MeV with width 0.655 MeV.

Theoretically, we find that $^5$He is unstable 
against neutron emission. The lowest state is the $3/2^-$ resonance calculated 
to be at 0.92 MeV energy, in good agreement with experiment. However, the 
$1/2^-$ resonance is predicted to be at a resonance energy of 4.47 MeV, at a 
much higher energy compared with the result of Ref.\cite{Til}. Some other 
theoretical investigations also predict a higher energy resonance for the 
$1/2^-$ state. For example, continuum shell model calculation of Volya and 
Zelevensky\cite{VZ} predicted the $3/2^-$ and the $1/2^-$ resonances at 0.99 
MeV and 4.93 MeV, respectively.

In $^{7}$He, the $3/2^-$ resonance is known to occur at a energy of 
approximately 0.45 MeV and has width $\Gamma=0.15$ 
MeV\cite{Wu08}.  A resonant state at $E^*$=2.9 MeV with width 
around $\Gamma$=2 MeV\cite{Wu08} was interpreted as odd nucleon 
coupled to the $2^+$ excited state of the core. Another resonant state at 
$E^*=0.6 (0.1)$ MeV with  $\Gamma$=0.75 MeV observed in the breakup of $^{8}$He
\cite{He74} was suggested to arise out of coupling of $p_{1/2}$ nucleon to 
the $^{6}$He ground state. Skaza \etal\cite{He75} observed  the 1/2$^-$ 
resonance at $E^*=0.9 (0.5)$ MeV with  $\Gamma$=1.0 MeV. Indications of a low 
energy narrow resonance was also observed by other workers\cite{He751}. 
However, a study\cite{He76} using isobaric analog states did not 
observe the last resonance but reported a broad 1/2$^-$ resonance at 2.2 MeV.
Ryezayeva \etal\cite{He77} also did not find any low energy 1/2$^-$ resonance 
but observed indications of a broad resonance at 1.45 MeV. 
Wuosmaa \etal\cite{He78,Wu08} observed a possible resonance at 2.6 
MeV excitation energy but 
no indication of any  1/2$^-$ resonance at lower energy.
Aksyutina \etal\cite{Aksyutina} found the resonance energy of the ground state to be 0.388 (2) MeV
and width 0.190 MeV. They could not draw any unambiguous conclusion about 
the possibility of a resonance around 1 MeV.

Our calculations place the $3/2^-$ resonance at 0.63 MeV energy. It appears at 
a higher energy possibly because the experimental state also has a contribution 
form $p_{1/2}$ orbit coupled to the $2^+$ state of $^6$He.
The $1/2^-$ 
resonance is  calculated to be at 2.09 MeV resonance energy, i.e. 1.46 MeV 
excitation energy. Thus our results agree with the experiment of Ryezayeva \etal
\cite{He77} and possibly with Skaza \etal\cite{He75} but not with other 
measurements. Other theoretical calculations 
also do not lead to any 
unambiguous conclusion. Continuum shell model study\cite{VZ} shows the 
ground state resonance 3/2$^-$ at 0.36 MeV and the excited 1/2$^-$ state around 
an excitation energy of 3.3 MeV. On the other hand, recoil corrected continuum 
shell model study of Halderson\cite{hal} place the resonance energy around 
1 MeV.

\section{Summary}

To summarize, structure of neutron rich He nuclei has been 
investigated using RMF approach in co-ordinate space.
The binding energy and radii values show reasonable agreement with experiment.
Optical model potentials have been calculated from effective interactions 
applied in finite nuclei in the folding model. Elastic partial scattering cross 
sections for proton scattering in inverse kinematics have been calculated using 
the theoretically obtained density for $^{6,8}$He and compared with some 
available experiments. The energies of the low-lying resonance states in the 
neutron unstable nuclei $^{5,7}$He have also been calculated and compared with 
experiments.

\section*{Acknowledgment}

This work was carried out with financial assistance of the
Board of Research in Nuclear Sciences, Department of Atomic Energy (Sanction
No. 2005/37/7/BRNS). One of the authors (MB) want to acknowledge the support 
by a grant from the Council of Scientific and Industrial Research, 
Government of India.


\clearpage

\end{document}